# Accused: How students respond to allegations of using ChatGPT on assessments[1]

Tim Gorichanaz
Drexel University
gorichanaz@drexel.edu

## Abstract

This study explores student responses to allegations of cheating using ChatGPT, a popular software platform that can be used to generate grammatical and broadly correct text on virtually any topic. Forty-nine posts and the ensuing discussions were collected from Reddit, an online discussion forum, in which students shared their experiences of being accused (the majority falsely) and discussed how to navigate their situations. A thematic analysis was conducted with this material, and five themes were discerned: a legalistic stance, involving argument strategy and evidence gathering; the societal role of higher education as a high-stakes gatekeeper; the vicissitudes of trust in students vs. technology; questions of what constitutes cheating; and the need to rethink assessment. The findings from this study will help instructors and institutions to create more meaningful assessments in the age of AI and develop guidelines for student use of ChatGPT and other AI tools.

## Introduction

The public launch of ChatGPT by the company OpenAI on November 30, 2022, took the world by storm. News of its capabilities spread through word of mouth and viral social media posts, and ChatGPT soon became the fastest-growing consumer web application in history, seeing 13 million unique visitors each day in January (Hu, 2023).

In technical terms, ChatGPT provides a conversational user interface for a large language model (LLM), in this case one called GPT (Generative Pre-trained Transformer), which was created through exposure to dozens of gigabytes of text (e.g., all of Wikipedia, millions of books, many websites such as Reddit) with human feedback during the training process.

Not long after its emergence, students and educators became aware of the capacities of ChatGPT for cheating. With plain-language input called a "prompt," ChatGPT responds with grammatical and often broadly correct plain-language output. *Write a five-paragraph essay*

---

*on the history of child labor in the United States at a college freshman level*, for example. Besides generating the essay all in one go, ChatGPT can generate outlines and thesis statements, rewrite text to provide more detail or strike a different tone, and cite sources. ChatGPT can also translate text, write code and more.

It sounds too good to be true, and it is. Though ChatGPT's output is often broadly correct, it is prone to generating misinformation—including making false statements, citing sources that do not exist and creating code that doesn't work—a propensity termed "hallucination" in the literature. Though OpenAI is working to address these hallucinations, some researchers argue that they are unavoidably inherent to LLM technology (Smith, 2023). Moreover, the very existence of systems such as ChatGPT is morally problematic, as they require enormous quantities of human-generated text as input, generally used without notice, consent or compensation, and they rely on underpaid human labor during the training process (Gorichanaz, 2023).

Despite all this, ChatGPT "kicked off an AI arms race," in the words of a *New York Times* headline (Roose, 2023), one that for now shows no signs of stopping. Competitors to OpenAI have released similar products and more are on the way. That term refers to the arms race between companies, but another arms race is unfolding between students and educators as both groups navigate what all this means for higher education.

The anxieties around ChatGPT in higher education center on academic integrity, learning and skill development, limitations of LLMs, policy and social concerns, and workforce challenges (Li et al., 2023; Sullivan et al., 2023). Meanwhile, ChatGPT is already prevalent among students. A survey by Study.com conducted in January 2023 found that about 90% of students used ChatGPT to help with homework and more than 50% used it to write an essay (Ward, 2023). Social media platforms such as TikTok are proving useful to students for sharing information about tactics for using ChatGPT to write essays and code and subvert detection methods (Haensch et al., 2023).

An early and predominant response has been the creation of so-called "AI detectors," systems meant to reveal whether a given text was LLM-generated. GPTZero, created by a college student, was publicized in early January 2023 (Bowman, 2023), later going on to raise millions of dollars in funding. A slew of competing AI detectors have emerged since then, including Originality AI, ZeroGPT, Writer AI Content Detector, OpenAI's own AI text classifier, and many others. Of particular note, Turnitin, makers of widely used plagiarism-detection software, released its AI detector in April 2023 (Knox, 2023).

But AI detectors, like LLM-generated content itself, are not reliable. They are prone to both false positives (human-generated text flagged as AI-generated) and false negatives (AI-generated text not flagged). And they can be gamed by using more precise prompts or by asking ChatGPT to rewrite the text, such as by specifying that it should use less expected language (Sadasivan et al., 2023; Wiggers, 2023).

Purveyors caution that AI detection should not be used as a standalone solution but only as one datapoint among many. The trouble is that bad information can be worse than no information.

Spring 2023 brought many news articles about instructors using AI detectors to punish students, sometimes wrongly. For instance, one professor failed his entire class after being told by ChatGPT that his students' work was AI-generated (Agomuoh, 2023). By June, even Turnitin admitted its AI detector wasn't reliable and should be used with caution (Fowler, 2023).

The present study seeks to help inform our understanding of how AI detectors are being used and interpreted, in this case by focusing on the student perspective, which Sullivan et al. (2023) found to be largely absent from news media discourse on ChatGPT and AI detection. In particular, the study examines how students experience and respond to accusations of using AI products in their written assignments. What is it like for students to be accused of cheating in this way? How do students make sense of the situation, and help each other do so? What strategies and tactics are shared for navigating the situation?

These questions are addressed through a thematic analysis of public posts made by anonymous students on Reddit, a discussion forum and one of the most visited websites on the internet. Dozens of such posts, with lengthy follow-up discussions, have been made in the months since ChatGPT's release, including both students who admitted to getting away with cheating and those who said they were falsely accused of doing so.

The findings from this study first and foremost help instructors, administrators and parents understand the student perspective in this situation. In doing so, the findings will help guide instructors in creating assignment instructions and guidelines for students and further conversation about the "AI arms race," particularly the one unfolding between students and faculty.

## Literature Review

To frame this study, the literature is reviewed on how educators respond to new technology, as well as cheating and strategies for addressing it.

### How Educators Respond to New Technology

One stereotype says that teachers don't want to adapt to new technologies, whether out of technophobia or inertia. Indeed, research findings about teachers' resistance to new technologies go back at least to 1920 (Hannafin & Savenye, 1993). Though some educators may seem tech-averse from the outside, Howard and Mozejko (2015) propose a different explanation. New technology should not be implemented for its own sake, they argue, and widespread calls for just that lead to teacher disengagement from these technologies, creating the impression that these teachers are anti-technology. Resonant with this argument, a study by Martin et al. (2020) showed that the most important factor that educators consider when deciding whether to adopt a new technology is its potential benefit to student learning.

Artificial intelligence is the emerging technology du jour, and calls are coming from all corners of society about leveraging it and limiting it, often in revolutionary and doomsaying tones. Teachers may feel forced to respond, yet the blurry and fast-moving picture makes it difficult to assess how these technologies will affect student learning. Risks to this situation have been voiced for several years, and they are encapsulated in a philosophical article by Guilherme (2019), who argues that in adopting new technologies, the relationship between students and teachers should not be overlooked. Emerging technologies, particularly AI-based ones, create a situation where teachers relate to their students more like objects than people. The justifications for implementing these technologies usually refer to efficiency, but the end goal of learning is not simply to learn more efficiently. Guilherme argues that technologies should be adopted in education only if they foster humane relationships rather than objectifying ones.

When it comes to some technologies, educators' hands may simply be forced. The dawn of the internet age—with widespread information access, distributed expertise and global connection—has unsettled the institution of higher education in many ways, cutting down to its very core. Laurillard (1999) warned that universities must either integrate these new technologies and evolve (e.g., to offer meaningful online learning experiences and flexible curricula) or fall back into small-scale elitism of their medieval form. "After some 25 years of fragmented experimentation with what computers might contribute to the teaching process, we still have rather little progress, or understanding of how to drive, rather than be driven by, the new learning technologies," Laurillard (1999, p. 135) wrote. "This would be nothing worse than a lost opportunity were it not for the fact that the new technology is also driving the world outside universities."

Many authors have echoed this sentiment. For example, Watty et al. (2016) position the uptake of new technology as the great challenge for higher education in the 21st century. McBride (2010) emphasizes that teachers need not act alone, but rather that the integration of new technology should be shepherded by university leadership and strategic planning, such as by investing in instructional designers and continuing training for instructors.

Looking more granularly at how specific technologies have been introduced in teaching and learning, Hartley (2007) provides a literature review and conceptual framework. According to Hartley, new technologies may be integrated into teaching in the following ways: direct instruction (in which technology replaces the human teacher); adjunct instruction (in which the teacher and technology work side by side); facilitating skills of learning (giving students practice with metacognitive skills); facilitating social skills (in which students work with each other alongside technology); and widening horizons (which allows learning to happen in new places and ways).

There are many examples in the literature of new technology successfully being implemented in the classroom to benefit students, including those reviewed by Hartley and others published more recently. Antonioli et al. (2014), for instance, report on introducing augmented reality in the classroom, showing that the technology taught students metacognitive skills, encouraged students to take responsibility for their learning, and was fun and engaging. But Antonioli et al. also observed challenges with budget and keeping the lessons learning-centered. Williams and Beam (2019) review the literature on new technologies in K–12 writing instruction, showing that new technologies improve students' writing while also supporting their higher-order thinking skills and also offer additional motivation to students that further drives learning—especially for reluctant writers. However, Williams and Beam point out that these new technologies suggest that the curriculum must adapt in certain ways, moving toward a process approach to writing (with iteration, feedback and collaboration, rather than the one-time submission of a finished product) and also engaging multimedia rather than only text.

Some work has focused specifically on chatbots and large language models (LLMs), which are especially relevant to ChatGPT. A recent meta-analysis suggests that AI chatbots can improve student learning outcomes, particularly in higher education (Wu & Yu, 2023); and a systematic review of LLMs in education shows that they can be useful for teachers in several tasks (generating questions, providing feedback, grading essays), but that these capabilities all raise practical and ethical concerns (Yan et al., 2023). Focusing specifically on ChatGPT for education, Su and Yang (2023) describe a number of benefits the tool may have, such as offering personalized and engaging learning experiences for students, and offering suggestions and efficient question-answering for teachers. However, Su and Yang acknowledge several barriers to these ends such as data quality and accuracy, cost, opacity of LLMs, and the inchoate state of LLM technology. Similarly, Kasneci et al. (2023) discuss numerous possibilities for LLMs to enhance learning, but raise additional challenges inherent in LLMs such as copyright issues regarding training data and output, the possibility of LLMs becoming a crutch for students or teachers, stakeholders' lack of expertise in the underlying technology, difficulties distinguishing the contributions of LLMs and students in submitted work, data privacy and security, sustainability and misinformation.

In the background of anxieties around new technologies are often concerns around student cheating. New technologies often mean new opportunities to cheat. Taylor (2003) relayed a story from the turn of the century, in which a teacher failed a large percentage of a class for plagiarism—resonant with a 2023 report mentioned in the introduction (see Agomuoh, 2023). Taylor writes that such reactions from instructors are a missed opportunity to teach students about what is and is not acceptable with new technology and to deepen their learning. More broadly, this work is a reminder that next technologies often stoke anxieties around cheating.

Building upon this literature, Rudolph et al. (2023) discuss the implications of ChatGPT for assessments in higher education, specifically writing assignments. As they write, faculty responses may include: going low-tech (requiring students to hand-write their work), using AI-detection tools (which are of dubious reliability), and creating assignments that AI cannot complete (such as responding to specific classroom discussions or referring to recent events or niche topics). But Rudolph et al. caution that these responses may only work in the short term and do not prepare students well for 21st-century society and employment. Better approaches, they write, include doing certain assessments in class, asking students to do oral presentations, moving from text-only essays to multimedia documents, letting students choose topics of genuine interest, using authentic assessments (that is, tasks that emulate realistic professional situations), and using teach-backs and peer evaluations.

## Understanding and Addressing Cheating

In an educational context, cheating is a category of behavior in which students get academic credit in a dishonest or deceptive way, such as copying another student's work, secretly using notes during an exam or plagiarizing text in an essay. The proliferation of digital technology has introduced new ways of cheating, and cheating is considered to be a bigger problem by educators, as assessments move online, where strategies to prevent cheating developed in face-to-face contexts are no longer relevant (Mellar et al., 2018; Moten et al., 2013; Williams, 2001). The major methods for cheating in today's world include plagiarism, collusion, file sharing, text spinning (using software to rewrite text), exam cheating, security breaching, and contract cheating (enlisting a third party to do an assessment) (Lancaster, 2022a).

Of these, contract cheating is perhaps the most relevant in a discussion of ChatGPT. Contract cheating is the purchase of custom-made work, typically from an "essay mill," to submit for an assessment. The research on contract cheating suggests that it is growing in prevalence; a systematic review by Newton (2018) reports that since 1978, 3.52% of students on average reported using contract cheating, but looking only since 2014 that number was 15.7%. A study of solicitations on Twitter found that students were willing to pay $33.32 per 1,000 words on average (Amigud & Lancaster, 2020). With the advent of ChatGPT and other readily available systems for generating text in response to a prompt, contract cheating becomes near-instantaneous and free. How essay mills and other institutions for contract cheating will respond will remain to be seen. (Note also that beyond replacing contract cheating, ChatGPT can be used for cheating in other ways, such as text spinning.)

Unfortunately, cheating appears to be common, with the majority of students reporting having cheated—many studies report over 75% (Baird, 1980; Crown & Spiller, 1998; Curtis, 2022). While there is evidence that cheating, particularly plagiarism, decreased over the period 1990–2020 (Curtis, 2022), other research suggests that cheating increased since the beginning of the Covid-19 pandemic (Jenkins et al., 2022).

Before going on, it's worth reflecting on why cheating is considered to be wrong. In a philosophical essay, Bouville (2009) addresses this question, finding that the oft-cited reasons for why cheating is wrong (it is unfair, it hinders learning) are not convincing. They betray inconsistencies and an underlying philosophy that school is about enforcing competition among students rather than education. In light of this, Bouville suggests that teachers concerned about cheating look deeper than the cheating behavior: "What hinders education is not cheating but the underlying lack of motivation: fighting cheating may only address a superficial symptom" (Bouville, 2009, p. 75).

## The Student Side of Cheating

To respond well to cheating, educators and institutions must understand its causes as deeply as possible. A large literature has investigated why students cheat. The reasons are numerous, including seating arrangements, knowledge of peer performance, high stakes for a given assessment (either reward or punishment), having failed before, having cheated before, having low or middling expectations for success, perceiving social norms that support cheating, perceiving assessments as unfair or irrelevant, low instructor vigilance, not having studied well, and dependence of financial support and long-term goals on good grades (Ahsan et al., 2022; Baird, 1980; Genereux & McLeod, 1995; Whitley, 1998). Regarding contract cheating specifically, the evidence suggests further that additional risk factors include speaking English as a learned language and being dissatisfied with the learning environment (Bretag et al., 2019). The research suggests that men are more likely to cheat than women, as are students with lower grades and those who believe the prevalence of cheating is high (Baird, 1980; Genereux & McLeod, 1995)—though a recent study suggests that women are more likely than men to cheat using digital technology (Krienert et al., 2021). When students who cheated were asked what would have stopped them from doing so, responses included more time, more resources, more skills to achieve the desired result, better time management, and less impact of mistakes on grades (Beasley, 2014).

The literature on student cheating also suggests that students do not always know what constitutes cheating (Beasley, 2014; Burrus et al., 2007; Raines et al., 2011). For example, Beasley (2014) reported that the majority of the students who cheated said they would not have done so if they knew what they were doing was cheating. A typical respondent in Beasley's study said, "If I knew what I was doing was wrong I wouldn't have done it plain and simple. ... I was unaware that my behavior was wrong" (Beasley, 2014, p. 235). Clarification here, according to Beasley's findings, would include clearer instructions, explanations of what constitutes plagiarism and how to avoid it (e.g., what "paraphrasing" means), a delineation of what behaviors are and are not acceptable (e.g., regarding collaboration on work), and explanation of the consequences for cheating. All this is even more important in settings that bring together students from multiple cultural backgrounds (Beasley, 2014). Instructors should also bear in mind that what constitutes cheating may differ from course to course (as learning goals differ) and across formats (norms for online learning are different from in-person learning) (Raines et al., 2011).

Regarding how students respond to being suspected of cheating, Pitt et al. (2020) report on student experiences of undergoing the formal disciplinary process after being suspected of contract cheating. In this study, some students did cheat and others did not. Across the board, the process was experienced as among the most challenging in the student's life, it created stress and vigilance around future assignments, it was kept as a secret to the extent that was possible, and it caused reputation damage with peers and faculty. All that said, these students were also able to guide other students toward practicing academic integrity.

## Instructor Responses to Cheating

Faced with cheating, instructors may address it or ignore it. Some studies suggest that upwards of 40% of instructors have ignored cheating when it was detected (Coren, 2011). The reasons for ignoring cheating include difficulties gathering convincing evidence, the time and effort it takes to follow the process, fear of retaliation or legal challenge, and the perceived triviality of the offense (Coren, 2011; Keith-Spiegel et al., 1998). Coren (2011) reports that instructors who had previous bad experiences with the student disciplinary process were more likely to ignore cheating.

Addressing cheating may be responsive or proleptic. Much of the literature on addressing cheating focuses on the proleptic—that is, taking measures to prevent future cheating. There are a number of approaches instructors may use for this, including education, technology, assessment design, sanctions, policy, and surveillance (Mellar et al., 2018). Given that new technologies have enabled new forms of cheating, many are turning to technology for solutions. The use of anti-plagiarism software has had some success (Ma et al., 2008) and may be partly responsible for a decrease in plagiarism in recent decades (Curtis, 2022). There is also evidence that contract cheating may be at least somewhat detectible with technology (Dawson & Sutherland-Smith, 2018; Lancaster, 2022b). But Mellar et al. (2018) emphasize that technology is not the primary means of addressing cheating but only one element among others.

There are numerous non-technological strategies for addressing cheating in the literature. In a review from the turn of the century, Williams (2001) discerned four key strategies: development of a culture of honesty, continual observation of student work, ongoing review of intermediate drafts, and face-to-face discussion about the work (Williams, 2001). In his book *Cheating Lessons*, Lang (2013) suggests emphasizing process over end product, implementing lower-stakes assessments, better preparing students for assessments, and fostering intrinsic motivation. In a recent systematic review, Ahsan et al. (2022) echo these suggestions, also adding that it is important to show students the institutional support that is available for issues they may be facing, and encouraging them to take initiative. Specific to preventing contract cheating, Ahsan et al. suggest crafting clear policies, providing support to students, and crafting assessments that dissuade students from using contract cheating. As technology changes, the academic environment needs to continually adapt in relation, when it comes to cheating. Teachers play a key but not the only role. (Lancaster, 2022a).

In the book *Cheating in College: Why Students Do It and What Educators Can Do About It*, the authors McCabe et al. (2012) particularly emphasize policy and culture. "We have a moral obligation to teach our students that it is possible and preferable to live and operate in an environment of trust and integrity where cheating is simply unacceptable" (McCabe et al., 2012, p. 165). This involves an educational component; sometimes students do not know or understand what constitutes cheating, especially when they receive mixed messages (e.g., about when collaborative work is allowed), so clear policies are vital. More broadly, the authors argue that students must perceive alignment between the formal systems and informal systems when it comes to academic integrity—from leadership and authority structure to mythos, norms and classroom mechanics. Doing so will require higher educational institutions to continue to reflect on the value and purpose of higher education (particularly as broader economic and political contexts change) and ensure that these are reflected in everything the institution does, from its academic integrity policies and how they are communicated, to the way online education is implemented (Rettinger & Gallant, 2022).

## Methodology

To analyze student experiences of and responses to accusations of using ChatGPT for cheating, data was collected from the online discussion website Reddit, one of the most visited websites on the internet and one often used for research of this nature. Reddit hosts several million discussion forums, called "subreddits," for countless topics, including college life and digital technology. Most Reddit users are pseudonymous, without any personally identifying information on their profile. In this anonymous forum, authentic conversations can proceed on sensitive issues, making Reddit an excellent source for data collection on this topic.

Data collection proceeded with searches on Google restricted to the domain "reddit.com" including terms such as "accused," "chatgpt," "essay," "professor," "AI," etc. An example of one of the complete queries used was: "site:reddit.com accused chatgpt professor." Searches were conducted periodically between May 21 and June 5, 2023. Resulting threads were included in the corpus if they were written by a college student (any posts by other students were excluded) directly involved in a case of AI writing allegation (rather than reporting on a friend's experience or commenting in general). In total, 49 threads were retrieved, details on which are given at the beginning of the Findings section below.

Analysis of these threads began with completing a spreadsheet to capture the high-level characteristics of each thread; the columns in this spreadsheet included: the date of the original post (the first post in a thread), the headline, the subreddit to which the thread belonged, a brief synopsis, whether the original poster (the user who started the thread) used AI in their work, the type of assignment, the emotional valence of the post, the AI detector mentioned, and the URL. After that, the contents of each thread were analyzed through reflexive thematic analysis (Braun & Clarke, 2012), an inductive form of qualitative analysis used to discern the major themes in a corpus. This analysis began with the open coding of salient quotes and proceeded through several rounds in which these codes gradually coalesced into themes.

Because the data used in this study were public and no personally identifiable information was collected or discernible in the corpus, this project was not deemed to be human subjects research by the researcher's institutional review board. Still, to protect the confidentiality of the people whose words contributed to the corpus in this study, details such as usernames, post headlines, URLs, etc., are not reported here; further, direct quotes given in this paper have been lightly edited ("disguised") to make them more difficult to locate via search as a means to protect their authors from harm, according to internet research best practices (Bruckman, 2002).

## Findings

The corpus for this study included 49 Reddit threads with posts ranging from December 21, 2022, to June 4, 2023, when data collection concluded. Of these threads, more than half appeared after May 1. The majority of the threads were from r/ChatGPT (n=25), followed by r/college (n=5), as well as several other subreddits with two or fewer posts each, including those for specific educational institutions.

The number of upvotes and comments on these threads ranged from less than 10 to over 2,000. The majority of original posts in the corpus had less than 50 upvotes and comments. Five had over 1,000 upvotes. The most upvoted thread saw over 45,000 upvotes and over 3,000 comments and was posted in r/mildlyinfuriating; following that, the next most upvoted thread had 15,000 upvotes and 2,500 comments and was posted in r/ChatGPT.

Within the corpus, 11 original posters said they had used ChatGPT on an assignment, while 38 said they were falsely accused of doing so. Of these 38, two mentioned they did use Grammarly. Most of the posts (n=43) described the type of assignment; most were essays (29), followed by discussion board posts (4). In all cases, the original posters were seeking advice for navigating their situation. Typically, an AI detector was used to ground the accusation. In 25 posts, a specific detector was mentioned; these included Turnitin (n=15), ChatGPT itself (6), and GPTZero (4). Five posts said the instructor had used multiple detectors. In most cases, the instructor requested a meeting with the student to discuss a high AI detection score; generally, students interpreted this request as an accusation. In some cases, instructors immediately reported a student to the student conduct office.

In terms of emotional valence, about half the posts were neutral in tone, with the rest expressing a strong emotion. These emotions included hostility, anxiety, fear and defeat. Respondents often offered support and condolences, particularly in the earliest threads; later on, some respondents mentioned that these sorts of posts were "getting old."

Besides offering advice and commentary, as will be discussed below, the content of many of these threads included technical discussions of how GPT works and why AI detectors are unreliable. A common refrain was that AI detectors are random number generators. In these discussions, certain misunderstandings about these issues were also evident. For example, several students and faculty used ChatGPT itself as an AI detector, asking the chatbot if it wrote a particular text. As one student shared, "One of my professors told us that a student used ChatGPT to write their essay and was promptly suspended. And all he had to do was ask ChatGPT if it wrote the essay. As a freshman that's TERRIFYING to me." But as other users pointed out, ChatGPT is not able to do the analysis required to make such an assessment.

Throughout these discussions, several themes were evident, which will be discussed below in turn.

## Legalistic Stance

Whether students were directly accused of cheating with ChatGPT or asked to an exploratory meeting to discuss the results of an AI detector, they seemed to experience the situation as a legal proceeding, and commentators further advised they treat it as such. A few threads discussed filing lawsuits, but in most cases the legalistic stance was more metaphorical. Along these lines, the following terms were common in the corpus: convince, evidence, logic, burden of proof, process, procedure, formal complaint, and disclaimer.

Much of the discussion centered on constructing arguments that would establish a student's innocence. One common tactic suggested was "deny, deny, deny," whether guilty or not, because the burden of proof is on the instructor and it is impossible to prove AI involvement in text generation except in very few cases (such as submitting work that includes phrases such as "As an AI language model," which ChatGPT uses in response to some prompts). Somewhat contrary to this, many felt that in the university setting a student is "guilty until proven innocent." Another common refrain in the corpus was "escalate," meaning to bring the issue to a higher authority than the instructor, such as the department head or dean. "Remember, you are the customer," one student wrote. "Escalate the situation until you get what you want." Students were advised to "get everything in writing," meaning to document every conversation and milestone as their case progressed.

The existence of AI detectors made argument construction all the more fraught. Some students asked why AI detectors flagged their work as AI-generated when they wrote it themselves. Some suggested ways to demonstrate that AI detectors are unreliable, while others felt that their very knowledge of AI detectors implicated them. "How do I explain why I put it through an AI detector if I'm not guilty?" one student wrote.

To ground these arguments, gathering evidence was suggested. Across the board, the best evidence was considered to be process materials from working on the assignment, such as notes, an outline and references, preferably written by hand. However, many students mentioned not having these materials. In such cases, the discussion often turned toward a "cover your ass" lesson for other students working on future assignments. To this end, enabling version history in Google Docs or Microsoft Word were also common suggestions; the version history would show that the text was not pasted in one step from ChatGPT. Some also counseled students to use a screen recorder when working and to share that recording should they be suspected of cheating.

Several acknowledged that these tactics verge on privacy invasion and should not be necessary, but conceded to using them because they had no other options. Others acknowledged that these tactics, in the end, are also limited. "It will work until they accuse you of using a second computer or your phone. There's no way to completely solve the problem." Others mentioned that with new features soon to be implemented by Google and Microsoft, word processing software will have ChatGPT-like features built in. "So the solution is to work with these tools, rather than trying to prove or disprove use. We have a lot of work to do to get educators, at all levels, up to speed."

Another stream of evidence related to establishing the unreliability of AI detectors. A common suggestion to this regard was to input some of the instructor's own writing into the detector, on the assumption that some of it will be flagged as likely AI-generated. Many shared that at least one detector claims that the Declaration of Independence is AI-generated. Some students suggested sharing news articles or a research paper about the accuracy of AI detectors.

It seems that some instructors, too, suggested that students check their work with AI detectors before submitting it. As one student said, "The professor has been trying to say if you run your work through the software and it gives a false positive, rewrite it until it does not say it's AI-generated." Reflecting on this, another student wrote, "Getting accused is seriously my worst fear. I've been pasting all my work into an AI detector. My own writing comes up as 'Likely AI.' It's stupid." While some students accepted this as the new status quo, others resisted: "No, don't let it go. AI detectors are a scam, and it's not our responsibility to adjust our writing to make them happy."

## The Societal Role of Higher Education

The next theme centered on the role of higher education as a gatekeeper for one's livelihood. At least in the United States, higher education is a major monetary investment, and many students perceive a college degree to be a prerequisite for gainful employment as an adult. As such, the stakes are perceived as very high in accusations of cheating. One student wrote, for example, that if universities are to use AI detectors, they need to be completely accurate and reliable. "There's no margin for error because the stakes are too high." Inevitably, a student's grades and GPA were wrapped up in these discussions.

Related to the central role of higher education, some students felt professors to be overly self-important in matters of cheating allegations. Comments along these lines were often sarcastic. For example: "They see themselves as The Authority. They aren't to be questioned, especially when they could be proven wrong." Another student wrote, "The idea of sucking up to these tyrants is sickening."

Given all this, several commentators expressed that they were glad to have graduated (or retired, in the case of instructors) before ChatGPT's release and therefore don't have to deal with this situation. One graduate wrote, "No one's gonna know what to do for a few years. I'm just glad I graduated already. I imagine there will be some things that are fair, others unfair. Some will skate by and graduate without lifting a finger during this time. Others will be expelled even though they didn't cheat."

In this situation, some saw a possibility for higher education to be unseated from its central gatekeeper role. Some remarked that higher education was now "irrelevant."

The centrality of higher education to society has another sense as well. Other students voiced that the issues unfolding in higher education are a harbinger of issues to come for society more broadly. "AI papers and people like you getting falsely punished for them, is just the tip of the iceberg. Society is screwed. A catastrophe is waiting to hit in the next few years." Issues such as political disinformation, polarization, the scientific enterprise, etc., are all susceptible to the same questions of authenticity as student assessments—and if the stakes for student work are high, the stakes for these things are much higher.

## Trust in Students and Technology

Another major theme in the corpus was trust—that between human and technology as well as among humans. The discussions in several threads reflect how trust can be built or damaged through new technology.

The central relationship in this theme was between instructor and student. Some students expressed that having built rapport with their instructor earlier in the term encouraged the instructor to give them the benefit of the doubt in a suspected case of AI cheating. Other students expressed surprise that even though they had felt a trusting relationship with their instructor, they still were accused. "I never would have expected to get accused by him, out of all my professors," one student wrote.

In either case, students said that an accusation of cheating damaged the relationship. For some, this meant they would have to work on future assignments defensively, expecting that they may be suspected or accused of cheating. "Screen recording is a good idea, since the teacher probably won't have as much trust from now on," one commentator wrote.

Trust was also evident in discussions of how instructors use AI detectors. "Of course she trusts the AI detector more than she trusts us," one student wrote.

Beyond the student–instructor relationship, trust was also discussed in student–student relationships, such as in work on group projects. A handful of threads shared experiences of a group submission being flagged as AI-generated, with students suspecting their groupmates of cheating. "I know I sure as hell didn't plagiarize but unfortunately you can't always trust others," one student wrote.

## What Constitutes Cheating?

The next theme reflects on the nature of cheating, which usages of AI constitute cheating (e.g., using AI to generate text vs. as a rephrasing tool), and whether AI technologies are already covered by existing intellectual integrity policies.

First, several students questioned why using AI was considered cheating. In these discussions, plagiarism was generally the concept invoked. For example, arguing that submitting ChatGPT's output could not be considered plagiarism, one user wrote, "Show your professor the ChatGPT terms of use. All rights to the generated content are assigned to you, to do with as you wish. You can't plagiarize what you own." Others pointed out that such a viewpoint overlooks some aspects of plagiarism, such as the possibility to self-plagiarize. Others suggested that even if using ChatGPT's output is not plagiarism, it is still academic dishonesty.

But some usages of ChatGPT were less clear. For example, one user asked, "Is using ChatGPT to rephrase parts of an essay with your own researched content considered cheating?" Considerations in this area were not just whether such usages would be considered cheating by human judgment, but also by a technological tool such as Turnitin. As one graduate student shared, "In some essays, I typed my own ideas out and used ChatGPT to refine and paraphrase them for me. Would this be considered plagiarism by Turnitin? I am freaking out."

Others noted confusion and contradiction in instructions they had previously received from faculty, especially when it comes to grammar-assistance tools lik Grammarly. "They used to recommend you use things like Grammarly, which use AI to correct your writing, and same with the grammar tools in Google Docs and Word, but now using those tools will get you flagged for AI-generated text." Some users mentioned they were now "scared" to use Grammarly.

Relatedly, some students noted hypocrisy in students being prohibited from using AI tools but not instructors. Several pointed out that the use of AI detectors are also AI tools. In one case, a student even thought their instructor was hypocritical with regard to plagiarism: "I had an online Zoom class where the teacher gave a 30-min speech about plagiarism, and then 2 weeks later said he had a video for us. He went on to play a screen recording from a different college's online Zoom class for the same subject. It's laziness and bluffing and I don't see it getting any better."

## Rethinking Assessment

In the final theme, conversations reflected a need to revise models for evaluating student learning. Some suggested that instructors would have to move toward oral or handwritten assignments done in the classroom to ensure that ChatGPT was not used; others remarked that certain types of assignments, such as online discussion posts, were overused and should be retired in favor of more authentic assessments.

Still, creating those assessments and operationalizing classroom policies was seen as a challenge. One student wrote of a professor who announced a policy that ChatGPT could be used for idea generation, rephrasing, etc., so long as the resulting Turnitin AI detector score was acceptably low. But that professor then penalized the student for submitting an annotated bibliography that scored too highly for AI content. The student felt betrayed. "I even ran her assignment descriptions through the OpenAI detector, and a few of them scored higher than even my assignments," that student wrote.

For some students, the impending change was welcome. "I can't wait for the entire college system to burn down over the summer because of AI," one student wrote. "Hopefully by fall, all of this will be cleared up."

In the meantime, ChatGPT allowed some students to spend less time on their peripheral courses and focus on the work they feel is more important. For example, one student wrote, "I'm a junior, and this was Sociology 101 where the teacher basically wanted us to echo her opinions about society and have no opinions. If I'm being told to lie, of course I'm going to find ways to make the work easier so I can focus on the 300- and 400-level courses that will matter more for my life and career."

## Discussion and Conclusion

The world is grappling with advances in AI technology, and higher education is no exception. This study has explored one aspect of this, how students respond to accusations of using ChatGPT to write essays, exam responses and other assessments. As discussed above, students take a legalistic stance to these accusations, gathering evidence and constructing arguments. Students' responses to these accusations demonstrate the high stakes of the situation and, more broadly, the societal role of higher education, as well as how new technologies impact trust. The shifting landscape of AI capabilities raises questions of what behaviors qualify as cheating and how to educate students and enforce policies with regard to cheating with AI, and they ultimately point to the need to rethink assessment for the age of AI.

The corpus used in this study demonstrates how being accused of cheating with AI can be harrowing for students, whether they cheated or not, just as Pitt et al. (2020) found regarding accusations of cheating generally. And it is notable that in the corpus examined here, the majority—78 percent—of the accused students were falsely accused.

At the heart of these accusations are AI detectors, suggesting hopefulness for a technological solution to the problems posed by generative AI for higher education. Given the proportion of false accusations in the corpus of this study, as well as recent press, it is clear that AI detectors are not the solution.

Even if an AI detector was highly accurate and reliable, it would be insufficient for its purpose. Consider the Copyleaks AI Content Detector, which claims a 0.02% false positive rate. In a university of 20,000 students, assuming four courses per student and five assignments per course, this very low false positive rate still suggests that 80 students per year will be falsely accused of cheating with AI. Moreover, the perceived infallibility of such tools may mean that innocent students are left unable to defend themselves.

Reliance on AI detectors as the primary means of addressing AI cheating also creates an environment where savvy cheaters can game the systems in place with relative ease. Already in the corpus examined here, techniques are being circulated to thwart the AI detectors on the market, such as using AI to rewrite its generated text with more syntactical change, randomized sentence length, and even inserted typos.

But beyond all this, AI detectors will only ever be able to detect AI-generated text, not AI-generated ideas about which students have written in their own words. Indeed, *The Chronicle of Higher Education* published an article in May 2023 by an undergraduate student with the headline "You Have No Idea How Much We're Using ChatGPT" (Terry, 2023), which made the point that ChatGPT can be used for intellectual tasks besides generating the text itself (e.g., generating a topic, thesis statement and supporting points), none of which can be found by an AI detector. So AI detectors are not a full solution for cheating with ChatGPT, perhaps not even a partial solution.

All this points to a need for better answers. In this study, the themes of the Societal Role of Higher Education and Rethinking Assessment offer some ideas in that direction. Clearly, a single term paper or exam submitted at the end of a course is no longer a valid assessment of a student's learning during that course. Rather than attempting to use AI detectors to evaluate whether these assessments are genuine, instructors may be better off designing different kinds of assessments: those that emphasize process over product, or more frequent, lower-stakes assessments. To this end, suggestions in the literature regarding teaching to dissuade and prevent cheating are just as relevant in the age of AI (e.g., Ahsan, 2022; Lancaster, 2022a; Lang, 2013; Williams, 2001).

Besides rethinking assessment, it is clear that instructors (or institutions) must establish clear policies on what usages of AI constitute cheating and why. While some usages of ChatGPT and similar tools could be considered forms of contract cheating or plagiarism, other usages are less clear. Implicitness and ambiguity are not helpful here. Further, instructors have an opportunity to educate students on the potential for wise use of these emerging AI tools, as they are likely to play a role in the professions for the foreseeable future. "Wise use" here may entail how the tools may be utilized, their limitations, and ethical considerations about their very existence.

This study has offered a look at how students experience and respond to allegations of cheating with ChatGPT. As a thematic analysis, the findings here should be taken as illustrative and indicative, not exhaustive or statistically generalizable. Moreover, the corpus in this study was limited to English-language text discussions on Reddit, which may not be fully generalizable to the college student population. Further research using other methods may be used to validate, extend or challenge these findings and provide a more comprehensive view.